# A Record-High Ion Storage Capacity of T-Graphene as Two-Dimensional Anode Material for Li-ion and Na-ion Batteries

Xiaoming Zhang,[a,b] Lei Jin,[b] Xuefang Dai,[b] Guifeng Chen,[b] and Guodong Liu[a,b]*

[a]State Key Laboratory of Reliability and Intelligence of Electrical Equipment, Hebei University of Technology, Tianjin 300130,
China

[b]School of Material Sciences and Engineering, Hebei University of Technology, Tianjin 300130, PR China.
E-mail:gdliu1978@126.com

**ABSTRACT:** Developing applicable two-dimensional (2D) electrode materials with high performance, especially with high ion storage capacity, has become an ever more obsessive quest in recent years. Based on first-principles calculations, we report that T-graphene, a new carbon-based 2D material, has a record-high Li/Na storage capacity. The capacity of T-graphene is as high as 2233.2 mA h g$^{-1}$ for Li, and can reach 2357.2 mA h g$^{-1}$ for Na, which are 6 times as much as that of the commercial graphite and are the highest among 2D anode materials identified so far. We demonstrate that the ultrahigh storage capacity of T-graphene mostly benefits from its low atomic mass and special periodic lattice structure. T-graphene has not only the ultrahigh storage capacity but also hosts the stable ion adsorption, good electric conductivity, fast ion diffusion speed, and low open-circuit voltage, which are merits required as a superior anode material for Li-ion and Na-ion batteries with ultrahigh storage capacity.

Keywords: Li-ion batteries; Na-ion batteries; Anode materials; T-graphene; Ultrahigh storage capacity.

# 1. Introduction

Rechargeable ion batteries, as one of the most representative clean energy-storage technologies, have received significant attention in current research [1,2]. Among them, Li-ion batteries (LIBs) gain the most success and have been broadly used in the daily life mainly due to their significant advantages of excellent reversible capacity, long cycle life, and high energy storage efficiency [3-6]. Nevertheless, with the increasing demand for high power electricity, it puts forward higher requirements of energy-storage capacity for LIBs. To enhance the capacity, one effective choice is to exploit excellent battery electrode materials, since the capability is closely related to the electrochemical characteristics of electrode materials [7-10]. Currently, the cathode materials for LIBs have been developed rapidly, however the choice of excellent anode materials is still limited. Graphite, as the most popular anode material, can show excellent cycling ability and has relatively low cost, but its storage capacity and rate ability are far from satisfactory [11,12]. Thus, in the current stage, it urgently needs to develop high performance anode materials, especially those which can host high storage capacity. Besides LIBs, it has also seen great research efforts on Na-ion batteries (NIBs) in these years [13-15]. This most benefits from the low cost and good operating safety of NIBs. Quite similar with LIBs, one major research focus on NIB technologies has also been cast in exploring high capacity anode materials.

Up to date, anodes based on two-dimensional (2D) materials have attracted increasing attention due to that the surfaces of 2D materials are fully exposed and expected to favor the ion diffusion and insertion processes [16-18]. Many 2D anode materials have been developed, such as graphene-based networks, MXenes, 2D metal nitrides and oxides [19-39]. Most of the proposed 2D anode materials for LIBs show storage capacity in the range of 400–800 mA h g$^{-1}$ [20-23,25-28,35], which are indeed much higher than the commercial graphite anodes (~372 mA h g$^{-1}$ [12]). Especially, the Li (and Na) capacity predicted by theory can reach 1984 mA h g$^{-1}$ on borophene [32], which is the highest among 2D anode materials reported so far. However, the efforts seeking for record-high ion-storage capacity in 2D anode materials have never diminished.

Carbon is a very "magic" element that can form extremely rich elemental crystals, ranging from the well-known graphite, diamond, graphene and C-60 fullerene [40,41], to the later reported M-carbon [42], V-carbon [43], bct-$C_4$ [44], and so on. Recently, a new carbon allotrope, namely T-carbon has been theoretically proposed and subsequently synthesized in experiments [45-47]. As the 2D form of T-Carbon, T-graphene has also been proved to be stable [48-50]. Gu *et al.* recently proposed a novel synthesis route to realize monolayer T-graphene, by using $C_4K$ compound as the medium [49]. Very interestingly, T-graphene was proposed to show many outstanding characters, such as Dirac-like fermions in the buckled form [48], excellent mechanical performance [50], and high critical-temperature superconductivity [49]. It is worth noticing that, T-graphene naturally has good electronic conductivity and very low elemental mass, which comfortably satisfy the requirements for high performance electrode materials. However, how T-graphene behaves as an electrode material has not been studied so far.

In this work, based on first-principles calculations, we make a thorough study on the performance of T-graphene as a potential electrode material for LIBs and NIBs. We find both Li and Na can achieve stable (chemical) adsorption on T-graphene with the metallic conductivity retained. The ion diffusion simulations suggest that T-graphene can offer excellent Li/Na diffusion with low diffusion barriers. Besides, T-graphene shows relatively low open circuit voltages for Li and Na, which is suitable to apply as battery anode material. Most remarkably, our results show that T-graphene can realize a record-high storage capacity for Li and Na among known 2D anode materials. The causes of such high Li and Na capacity in T-graphene will also be demonstrated.

## 2. Computational details

Our calculations are performed by using the Vienna ab initio simulation package (VASP) [51], on the basis of density functional theory (DFT) [52]. The exchange-correlation is treated as the generalized gradient approximation (GGA) of Perdew−Burke−Ernzerhof (PBE) functional [53,54]. The cutoff energy is chosen as 400 eV during the calculations. For T-graphene monolayer, we built a vacuum with the thickness of 18 Å to avoid potential interactions between layers. During calculations, the long-range van der Waals interactions are taken into account by using the DFT-D2 method [55]. To study the ion adsorption and diffusion processes, a 3 × 3 supercell of T-graphene monolayer is applied, with the atomic positions are fully optimized. The k-mesh is chosen as 7 × 7× 1 and 9 × 9× 1 during the geometrical optimization and static energy calculations. The force and energy convergence criteria are set as 0.01 eV Å$^{-1}$ and $10^{-8}$ eV, respectively. The Bader charge method [56] is applied to evaluate the charge transfers during the ion adsorption process. The climbing-image nudged elastic band (CI-NEB) method [57,58] is used to obtained the diffusion barrier height during the ion diffusion process. The phonon spectra are calculated by using the PHONOPY package [59,60].

## 3. Results and discussions

### 3.1 Crystal structure, stability and electronic structure of T-graphene

Figure 1(a) shows the 3 × 3 supercell of T-graphene monolayer. The C-C bonding in T-graphene forms two geometrical configurations: one is tetragon, and the other is octagon. The octagon is constructed by two different bonds with the bond lengths of 1.48 Å (inter-tetragon) and 1.35 Å (intra-tetragon), respectively. The shadow region in Fig. 1(a) shows the primitive cell of T-graphene monolayer. Its crystal structure belongs to the plane group *p4mm* and each primitive cell contains four C atoms. After geometry optimization, the lattice constant yields to be 3.45 Å, in good accordance with former results [48,49]. The formation energy of T-graphene is calculated to be -8.78 eV per atom, being comparable with well-known carbon allotropes including graphene (-9.26 eV), bct-$C_4$ (-8.92 eV), and graphdiyne (-8.49 eV) [45,50,61], which suggests good thermodynamic stability of T-graphene. We have further checked the dynamic stability of T-graphene from the phonon spectra. As shown in Fig. 1(c), it exhibits no imaginary modes in the phonon spectra, indicating excellent dynamic stability of T-graphene. Unlike the insulating T-carbon, T-graphene shows metallic band structure. As shown in Fig. 1(d), it can be observed that there exist two bands crossing the Fermi level. The obvious metallic conductivity makes T-graphene potential to be applied as electrode material for ion batteries.

**3.2 Ion adsorption and diffusion processes on T-graphene**

Examining the geometrical symmetry of T-graphene, there typically possess five potential ion adsorption sites (S1-S5), as shown in Fig. 1(b). To determine the most stable Li/Na adsorption site, we compare the adsorption energies for single Li/Na atom on these sites. The adsorption energy can be described as:

$$E_{Ad} = E_{Li/Na@TG} - E_{TG} - E_{Li/Na} \qquad (1)$$

Here, $E_{Ad}$ is known as the adsorption energy; $E_{Li/Na@TG}$ and $E_{TG}$ denote the total energies of T-graphene supercell (3 × 3) with and without Li/Na adsorptions; and $E_{Li/Na}$ is for energy per atom in bulk Li/Na metals. Our calculations show that all the adsorption sites have the negative adsorption energies, indicating that Li/Na can be adsorbed on T-graphene. Among the five sites, we find S1 site possesses the lowest adsorption energy for both Li (-1.155 eV/atom) and Na (-1.134 eV/atom). The S2, S3 and S4 sites possess the same adsorption energy (-0.713 eV/atom for Li; -0.701 eV/atom for Na). These results suggest that Li/Na atoms the most prefer to be adsorbed on the center of the carbon-octagon (S1 site). For the optimized adsorption structure, we find adsorption height for Li is about 1.370 Å above the T-graphene, and that for Na is 1.872 Å.

To get further insights on the adsorption process, we have calculated the charge density difference (CDD) for the optimized Li/Na adsorption. As visualized by the CDD maps in Fig. 2(a), we can clearly observe that a large amount of charge transfers to the T-graphene substrate from Li/Na atom. The specific amount of transferred charge can be evaluated by the Bader charge calculations [56]. As the results, we find the transferred charge is 0.89 $e$/atom for Li, and 0.87 $e$/atom for Na, respectively. The CDD maps and the Bader charge analysis unambiguously suggest that Li/Na atoms are chemically adsorbed on T-graphene. Such adsorption process corresponds to the redox reaction for electrode materials. Figure 2(b) display the total density of states (DOS) for T-graphene before and after Li/Na adsorption. We clearly find that T-graphene retains good conductivity with metallic electronic structure after adsorption, which further ensures its feasibility to apply as a battery electrode material.

The rate performance is another key parameter for battery electrodes. Here we evaluate the Li/Na rate performance on T-graphene by performing CI-NEB calculations [57,58], as it has been proved effective to determine optimal ion diffusion path and the minimum diffusion barrier [20-39]. Three typical ion diffusion paths [Path-1, Path-2, and Path-3 in Fig. 3(a)] need to be considered in T-graphene. The calculated diffusion profiles for these cases are provided in Fig. 3(b). We can observe that Path-2 severs as the optimized Li/Na diffusion path. The minimum diffusion barrier is 0.44 eV for Li and 0.41 eV for Na. To be noted, the Li diffusion barrier on T-graphene is higher than many typical 2D anode materials, such as $VS_2$ (0.22 eV [20]), $MoS_2$ (0.21 eV [26]), and some MXenes (0.02-0.28 eV [10,20-22,26]), but comparable with graphene (0.37 eV [19]), and lower than the commercial anode graphite (~ 0.6 eV [12,62,63]). This suggests that T-graphene can still offer good rate performance for Li ion. We also notice that, the diffusion barrier for Na is a bit lower than Li, indicating T-graphene offers better rate performance for Na.

### 3.3 Record-high Li/Na storage capacity on T-graphene

The maximum ion storage capacity is known as the most crucial parameter for electrode materials, which can be obtained by evaluating the layer-resolved adsorption energies. We still use the 3 × 3 supercell of T-graphene as the substrate, then gradually increasing adsorption layers on both sides of the substrate. Based on the adsorption energies calculated above, the adsorption sequence should be: S1 (1$^{st}$ layer)-S4 (2$^{nd}$ layer)-S5 (3$^{rd}$ layer). The layer-resolved adsorption energy [$E_{Ad(layer)}$] has the following form:

$$E_{Ad(layer)} = (E_{Li_{18n}/Na_{18n}C_{36}} - E_{Li_{18(n-1)}/Na_{18(n-1)}C_{36}} - 18E_{Li/Na})/18 \quad (2)$$

where $E_{Li_{18n}/Na_{18n}C_{36}}$ and $E_{Li_{18(n-1)}/Na_{18(n-1)}C_{36}}$ denote the total energies of T-graphene supercell (3 × 3) with adsorbing $n$ and $n-1$ Li/Na layers, and the number "18" indicates eighteen adatoms for each adsorption layer.

For T-graphene, the common half-cell reaction has the form:

$$\text{T-graphene} + x\text{Li}^+/\text{Na}^+ + xe^- \longleftrightarrow \text{Li}_x/\text{Na}_x@\text{T-graphene} \quad (3)$$

Accordingly, the average open-circuit voltage ($V_a$) can be described as:

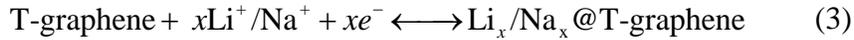

$$V_a = (E_{C_{36}} + xE_{Li/Na} - E_{Li_x/Na_xC_{36}})/xe \quad (4)$$

where $E_{C_{36}}$ ($E_{Li_x/Na_xC_{36}}$) represent the total energies of T-graphene before (after) Li/Na adsorptions, and $x$ represents the number of adatoms on T-graphene. Thereby, the specific voltage ($V_a$) versus the adatom concentration ($x$), known as the voltage profile, can well describe the adsorption sequence and voltage evolution during the layer-resolved adsorption. For Li and Na adsorptions on T-graphene, the voltage profiles are provided in Fig. 4(a) and (b), respectively. In the voltage profiles, each plateau corresponds to one layer adsorption. For Li adsorption, as shown in Fig. 4(a), the voltage keeps positive for the 1$^{st}$ and the 2$^{nd}$ plateaus, which indicates that at least two layers of Li can be adsorbed on T-graphene. During the adsorption, the voltage ranges in 0.18-0.08 V. For Na adsorption, as shown in Fig. 4(b), the voltage can also retain positive for the 1st and the 2nd layer adsorption. For the 3rd layer, we find at least two Na atoms can be adsorbed on the T-graphene supercell, which produces the short plateau above the zero voltage. During this whole period, the voltage change in the range of 0.44-0.02 V for Na adsorption. These low voltage values indicate that T-graphene is suitable to apply as anode material for LIBs and NIBs. The voltages in T-graphene are comparable with other 2D anode materials such as borophene (0.12 V to 0.68 V), GaN monolayer (0.05 V to 0.26 V), $Mo_2C$ monolayer (~ 0.14 V), 2D electride $Ca_2N$ (0.09 V to 0.23 V), and so on [20-22,31,32,64-67].

From the voltage profiles, we can find that the accommodated numbers of Li and Na atoms on T-graphene 3 × 3 supercell can be 36 and 38 at least. The corresponding chemical stoichiometries are of $Li_{36}C_{36}$ and $Na_{38}C_{38}$, respectively. Then, the ion capacity ($C$) on T-graphene can be obtained by following the equation:

$$C = x_m F / M_{T\text{-}G} \qquad (5)$$

Here, $x_m$ is the ion concentration (36 for Li, 38 for Na) on T-graphene supercell, $F$ is known as the Faraday constant (26.8 A h $mol^{-1}$), and $M_{T\text{-}G}$ represents the mass of T-graphene supercell ($C_{36}$). As the results, the Li storage capacity on T-graphene is calculated to be 2233.2 mA h $g^{-1}$, and that for Na is 2357.2 mA h $g^{-1}$.

Here we want to comment on the ultrahigh ion storage capacity of T-graphene, which can reach 2233.2 mA h $g^{-1}$ for Li and 2357.2 mA h $g^{-1}$ for Na. In Fig. 5(a) and (b), we compare the storage capacity between T-graphene and representative 2D anode materials for LIBs and NIBs. As shown in Fig. 5(a), most of the 2D LIB anode materials host the Li capacity lower than 600 mA h $g^{-1}$, such as $MoS_2$ [26], $MoN_2$ [31] and some MXenes including $Mo_2C$, $Ti_3C_2$, and $Nb_2C$ [20-22,27,28,64]. The Li capacity in T-graphene can be 3-7 times higher than these anode materials. Several 2D anode materials such as $Mn_2C$ [10], black-phosphorus, $GeP_3$ [65] and 2D GaN [66] have Li capacity in the range of

600-1000 mA h g$^{-1}$, which are still 2-4 times lower than that of T-graphene. Some 2D anode materials including monolayer W [67], $\beta_{12}$ and $\chi_3$ borophene [32], show extremely high Li capacity (> 1200 mA h g$^{-1}$). Among them, $\beta_{12}$ borophene show the highest Li capacity (1984 mA h g$^{-1}$ [32]). However, the capacity in T-graphene can be even 249 mA h g$^{-1}$ higher than $\beta_{12}$ borophene. T-graphene has reached a new record of Li capacity in 2D anode materials. Besides, we also show the data of commercial graphite [62,63] for comparison in Fig. 5(a), where the Li capacity in T-graphene can be 6 times of graphite (2233 mA h g$^{-1}$ *vs.* 372 mA h g$^{-1}$). For 2D NIB anode materials, as shown in Fig. 5(b), the Na capacity in T-graphene (2357.2 mA h g$^{-1}$) is also the highest among all the known 2D anode materials.

Before closing, we have two additional remarks. Firstly, monolayer form of T-graphene has not been experimentally synthesized currently. However, very fortunately, a recent work by Gu *et al.* [49] has proposed a promising synthesis route to obtain T-graphene monolayer. The synthesis route follows three processes: (1) synthesize the intercalation compound $C_4K$ from T-graphene and potassium at high pressure; (2) quench the intercalation compound into ambient condition; (3) exfoliate the T-graphene monolayer from $C_4K$ by available methods. One can follow this synthesis route to prepare T-graphene and further investigate its performances as battery electrodes experimentally.

Secondly, we want to point out that the record-high ion capacity in T-graphene mostly benefits from three key factors: (1) the carbon element has very low atomic mass; (2) there exist a large amount of periodic voids in the T-graphene lattice, which has increased stable ion adsorption sites and meanwhile lowered the mass density; (3) T-graphene can offer multilayer ion adsorption (2 layers for Li and 3layers for Na), which greatly raises the storage capacity.

## 4 Summary

In summary, we systematically investigate the feasibility of T-graphene as an electrode material for LIBs and NIBs by first-principles calculations. Our results show that both Li and Na prefer to adsorb on the hollow sites of T-graphene. The adsorptions are chemically stable with sufficient charge exchanges between the adatoms and substrate. During adsorption, T-graphene retains metallic, which ensures good conductivity as an electrode material. The open-circuit voltages are calculated to be as low as 0.08 V for Li and 0.02 V for Na, which suggests T-graphene is a promising LIB/NIB anode material. Moreover, T-graphene can offer excellent rate performance for Li and Na with the diffusion barriers lower than the commercial anode material graphite. Most excitingly, we find the Li storage capacity in T-graphene can be as high as 2233.2 mA h $g^{-1}$, and that for Na is 2357.2 mA h $g^{-1}$. These values are 6 times of commercial graphite anode, and are the highest among the 2D anode materials reported so far. Our results highly suggest that T-graphene is very promising to be developed as a super anode material for LIBs and NIBs with offering ultrahigh ion storage capacity.

**Acknowledgements**

This work is supported by the Nature Science Foundation of Hebei Province (Nos. E2019202222 and E2019202107). One of the authors (X.M. Zhang) acknowledges the financial support from Young Elite Scientists Sponsorship Program by Tianjin.

**Figures and captions:**

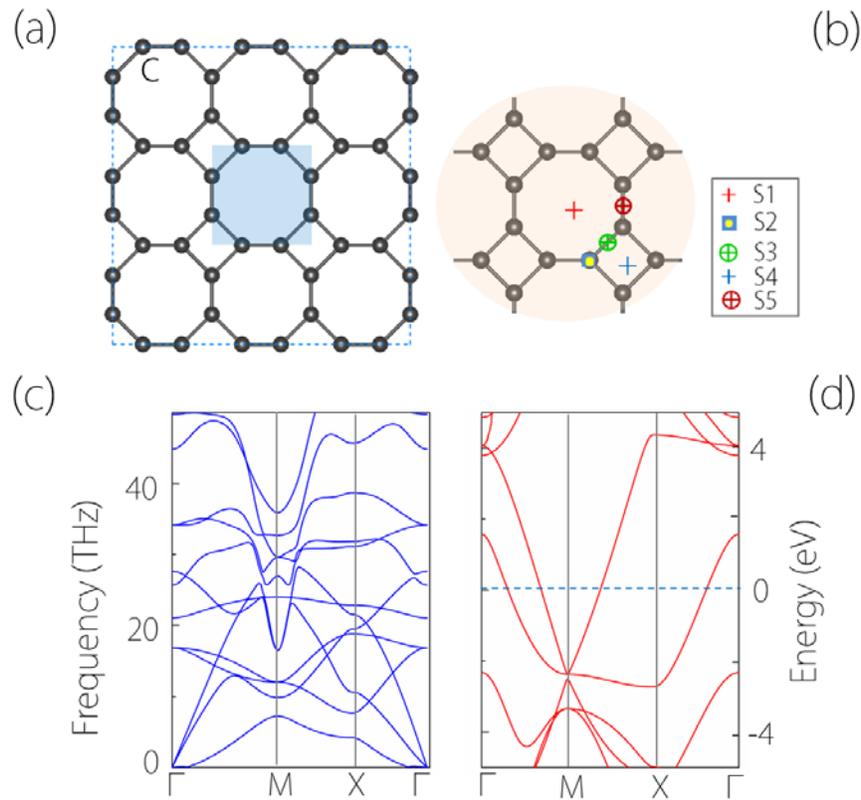

**Fig. 1** (a) Crystal structures of T-graphene in the form of 3 × 3 supercell. The shadowed region shows its primitive cell. (b) Potential Li/Na adsorption sites on T-graphene. The five sites are denoted as S1, S2, S3, S4, and S5, respectively. (c) The phonon spectra and (d) the band structure of T-graphene.

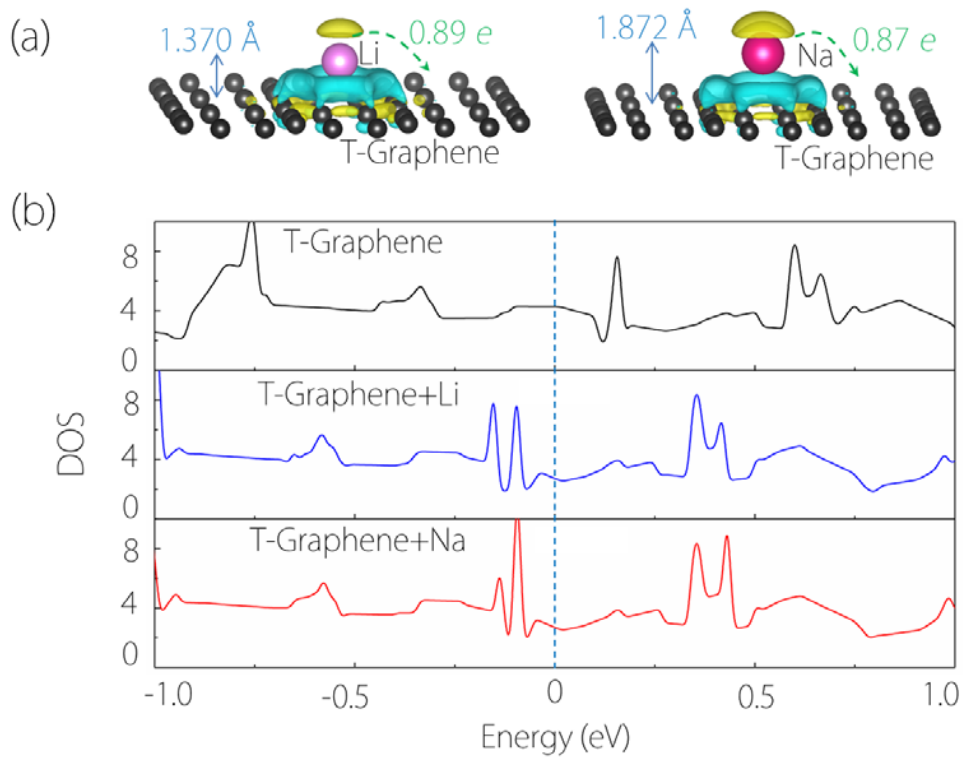

**Fig. 2** (a) Charge density differences of T-graphene after Li and Na adsorption. The regions with green and yellow colors indicate the electron accumulation and depletion, respectively. In (a), the amounts of transferred charges and adsorption heights are provided. (b) Total density of states of T-graphene before and after Li/Na adsorptions.

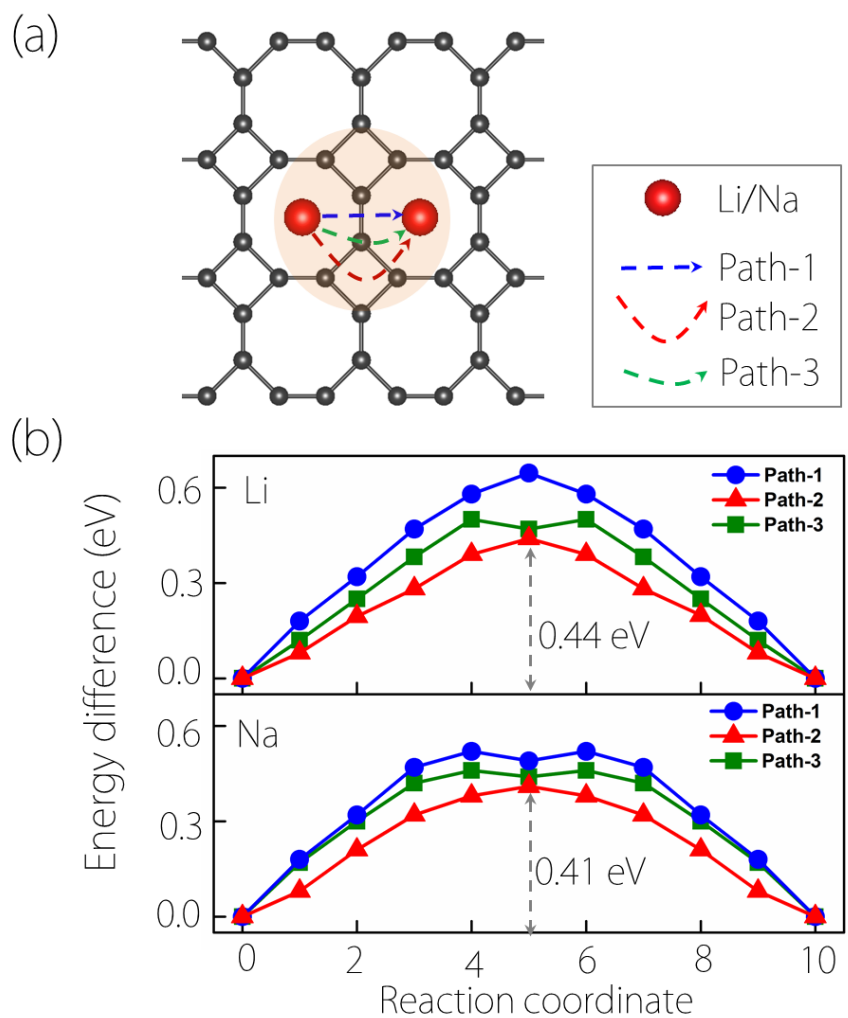

**Fig. 3** (a) Potential Li/Na ion diffusion paths and (b) corresponding ion diffusion profiles on T-graphene. In (a) and (b), the minimum diffusion barrier is shown.

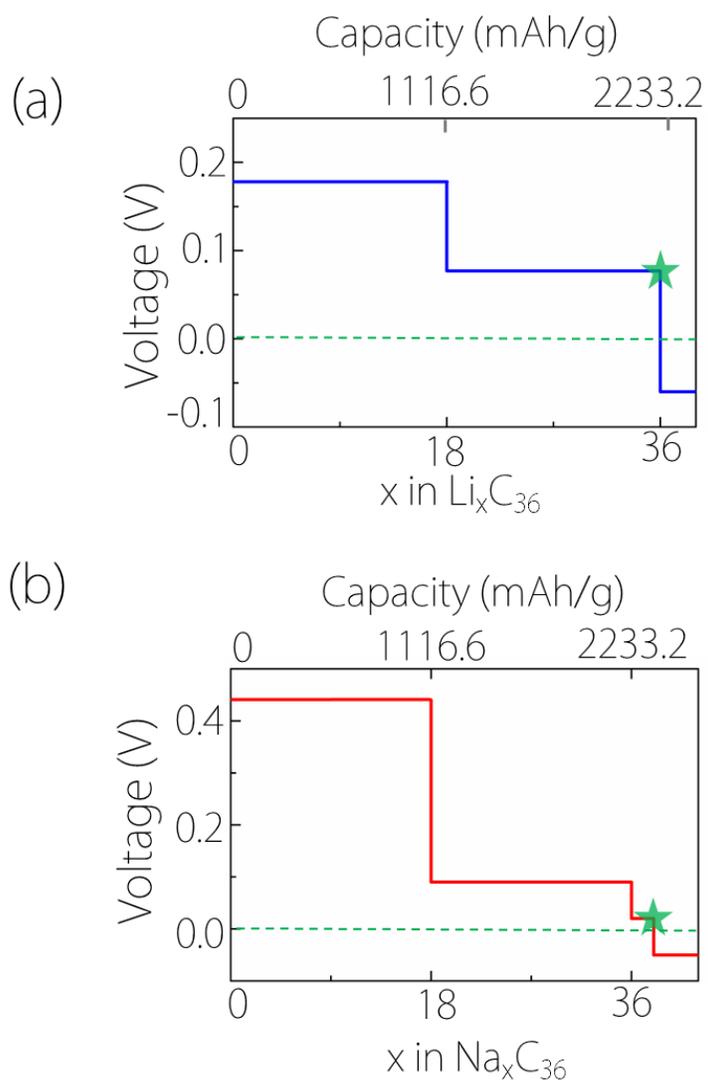

**Fig. 4** The voltage profiles and storage capacities for (a) Li and (b) Na on T-graphene. The positions of maximum storage capacities are indicated by the stars in the figures.

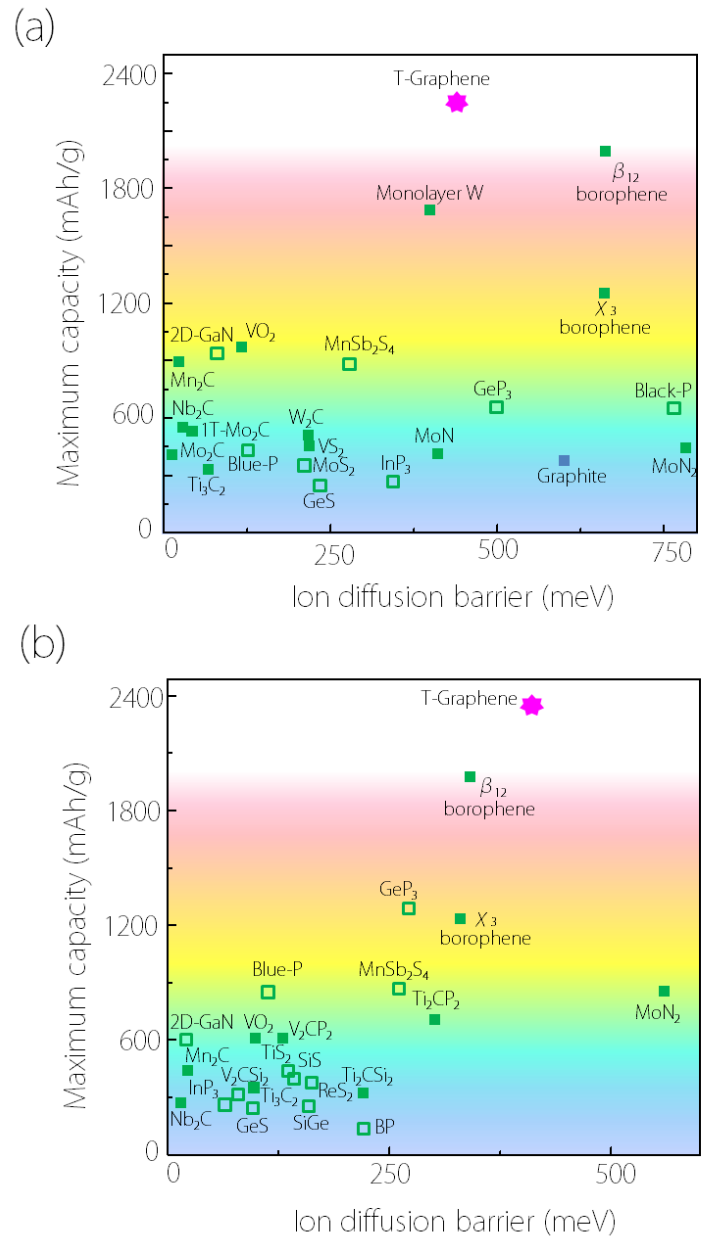

**Fig. 5** Comparison of the maximum ion capacity between T-graphene and typical 2D anode materials, where (a) is for Li and (b) is for Na. In (a) and (b), the metallic (insulating) anodes are differed by the solid and hollow squares. Some data are picked from literatures [19-39,64-74].